\documentclass[runningheads]{llncs}
\usepackage[T1]{fontenc}
\usepackage{amsmath}
\usepackage{multirow}
\usepackage{graphicx,verbatim}
\usepackage{url}
\usepackage{hyperref}
\begin{document}
\title{From Variability To Accuracy: Conditional Bernoulli Diffusion Models with Consensus-Driven Correction for Thin Structure Segmentation}
\titlerunning{Diffusion Segmentation and Consensus-Driven Correction}
\author{Jinseo An\inst{1} \and
Min Jin Lee\inst{1} \and
Kyu Won Shim\inst{2} \and
Helen Hong\inst{1}\thanks{Corresponding author}}
\authorrunning{J. An et al.}
%
\institute{Department of Software Convergence, Seoul Women's University, Seoul, Republic of Korea \\
\email{\{jsan,minjin,hlhong\}@swu.ac.kr} \and
Department of Pediatric Neurosurgery, Craniofacial Reforming and Reconstruction Clinic, Yonsei University College of Medicine, Seoul, Republic of Korea\\
\email{shimkyuwon@yuhs.ac}}
    
\maketitle              

\begin{abstract}
Accurate segmentation of orbital bones in facial computed tomography (CT) images is essential for the creation of customized implants for reconstruction of defected orbital bones, particularly challenging due to the ambiguous boundaries and thin structures such as the orbital medial wall and orbital floor. In these ambiguous regions, existing segmentation approaches often output disconnected or under-segmented results. We propose a novel framework that corrects segmentation results by leveraging consensus from multiple diffusion model outputs. Our approach employs a conditional Bernoulli diffusion model trained on diverse annotation patterns per image to generate multiple plausible segmentations, followed by a consensus-driven correction that incorporates position proximity, consensus level, and gradient direction similarity to correct challenging regions. Experimental results demonstrate that our method outperforms existing methods, significantly improving recall in ambiguous regions while preserving the continuity of thin structures. Furthermore, our method automates the manual process of segmentation result correction and can be applied to image-guided surgical planning and surgery.


\end{abstract}
\section{Introduction}

Orbital bone fractures commonly occur in thin regions, such as orbital medial wall and orbital floor~\cite{cho2013combined}. Accurate orbital bone segmentation in computed tomography (CT) images is crucial in craniomaxillofacial surgery, particularly for designing patient-specific implants and establishing image-guided surgical plans. However, segmenting these thin bone structures presents significant challenges due to their low contrast with surrounding tissues and ambiguous boundaries caused by partial volume effects in thin structures~\cite{kim2019three}, leading to inter-observer variability in manual annotations. Previous study has quantified this variability using intra-class correlation coefficient (ICC) in manual orbital bone annotation, reporting lower consensus in thin bone regions (ICC=0.715 for orbital medial wall, ICC=0.824 for orbital floor) compared to whole orbital bone (ICC=0.931)~\cite{kim2022comparative_3dpm}. Despite recent attempts such as MSDA-Net~\cite{an2023orbital}, which applied multi-scale and dual attention modules to improve segmentation accuracy for orbital bones of varying thickness, evaluation results showed variation depending on which annotation was used as reference standard~\cite{an2023evaluation}.

Recently, diffusion models have shown remarkable advances in medical segmentation tasks~\cite{yan2024cold,chowdary2023diffusion}. Unlike traditional CNN-based methods that produce deterministic results, diffusion models leverage the stochastic nature of noise sampling to generate diverse plausible segmentations. We argue that this inherent capability can provide insight when dealing with particularly thin structures and ambiguous boundaries that even expert annotators exhibit significant variability. Additionally, conditional diffusion models improve segmentation performance by incorporating anatomical structure in-formation from medical images as conditioning input~\cite{amit2021segdiff,rahman2023ambiguous}. While most existing studies rely on Gaussian noise, several studies have proposed using Bernoulli noise instead, which is more appropriate for binary mask segmentation tasks due to the discrete nature of the masks~\cite{chen2024hidiff,wang2023binary}. 

In this paper, we present a novel framework that combines diffusion model-based segmentation with consensus-driven correction to improve accuracy in challenging regions. Our main contributions are: (1) We employ a conditional Bernoulli diffusion model for segmentation, providing three different annotation masks per input image to learn inter-observer variability. (2) We propose consensus-driven correction to address the inherent variation in ambiguous boundaries, considering position proximity, consensus level similarity, and gradient direction similarity across multiple segmentations from the diffusion model. (3) Experimental results on thin bones of orbital medial wall and orbital floor demonstrate that our method corrects challenging regions and segmentation performance outperforms other comparison methods.

\section{Method}
\subsection{Conditional Bernoulli Diffusion Model for Image Segmentation}

\begin{figure}
\includegraphics[width=\textwidth]{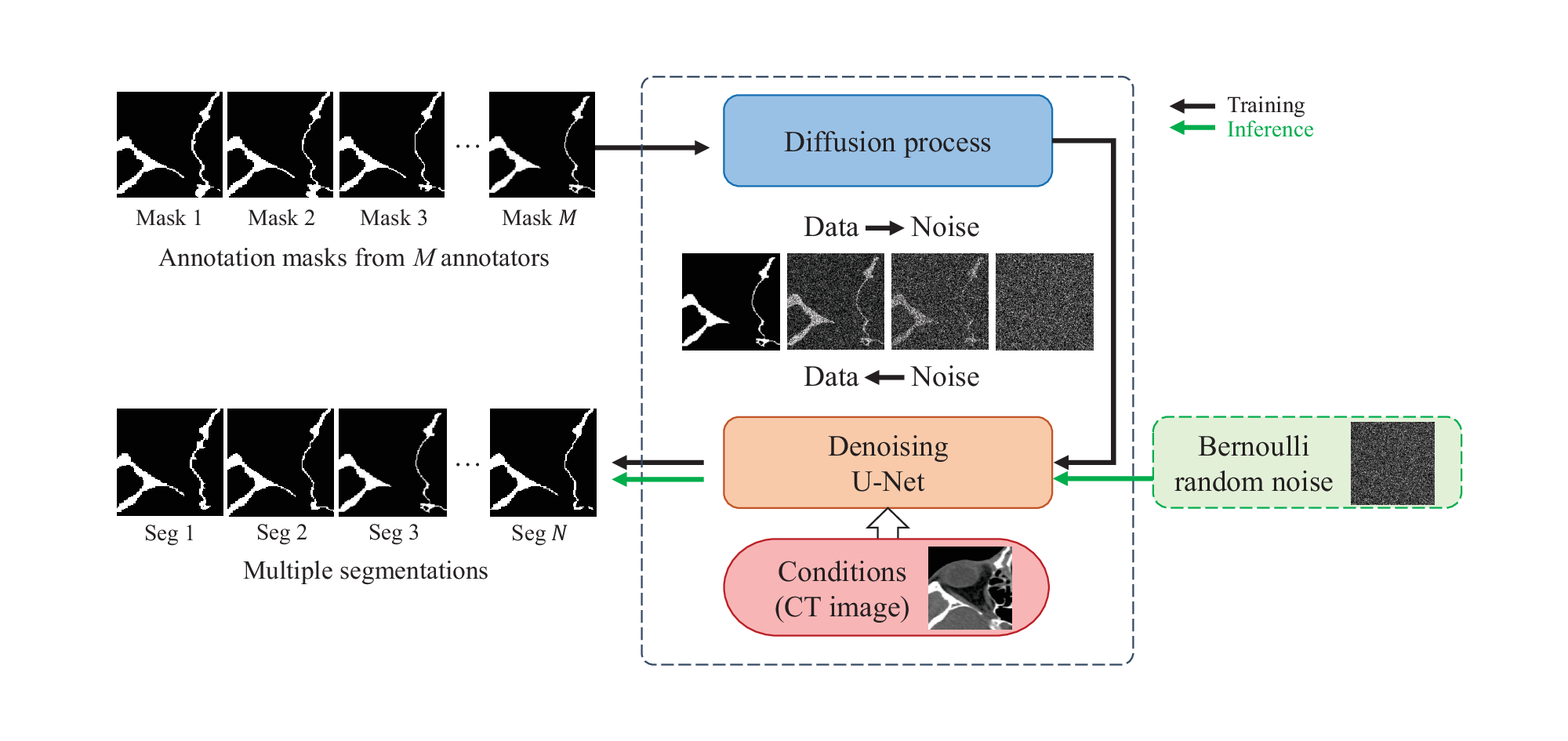}
\caption{Illustration of our conditional Bernoulli diffusion model} \label{fig1}
\end{figure}

Fig.~\ref{fig1} presents the overview of our conditional Bernoulli Diffusion Model for learning diverse annotation patterns and generating multiple plausible segmentations. While previous diffusion models typically employ Gaussian noise, we utilize Bernoulli noise-based diffusion model~\cite{chen2023berdiff} as more appropriate for binary medical image segmentation tasks, where masks consist exclusively of discrete values (0 or 1).

The model takes $M$ annotations from multiple annotators per input image as that capture inter-observer variability. During the forward process, binary masks adds random noise through Bernoulli noise sampling. In the reverse process, the model progressively removes noise to generate segmentations, conditioned on the corresponding CT images. The CT images used as conditions provide intensity and context information, enabling the model to understand anatomical structures during the denoising process.

An advantage of our approach is the ability to learn from various annotation patterns, reducing dependency on subjective annotation by any single annotator. By training on $M$ different annotations per input image, the model captures the inherent inter-observer variability in challenging regions. This is particularly valuable for ambiguous boundaries where even expert annotators disagree. In these regions, our diffusion model generates varying plausible segmentations based on the learned annotation distribution, which we identify as challenging areas requiring consensus-driven correction.

\begin{figure}
\includegraphics[width=\textwidth]{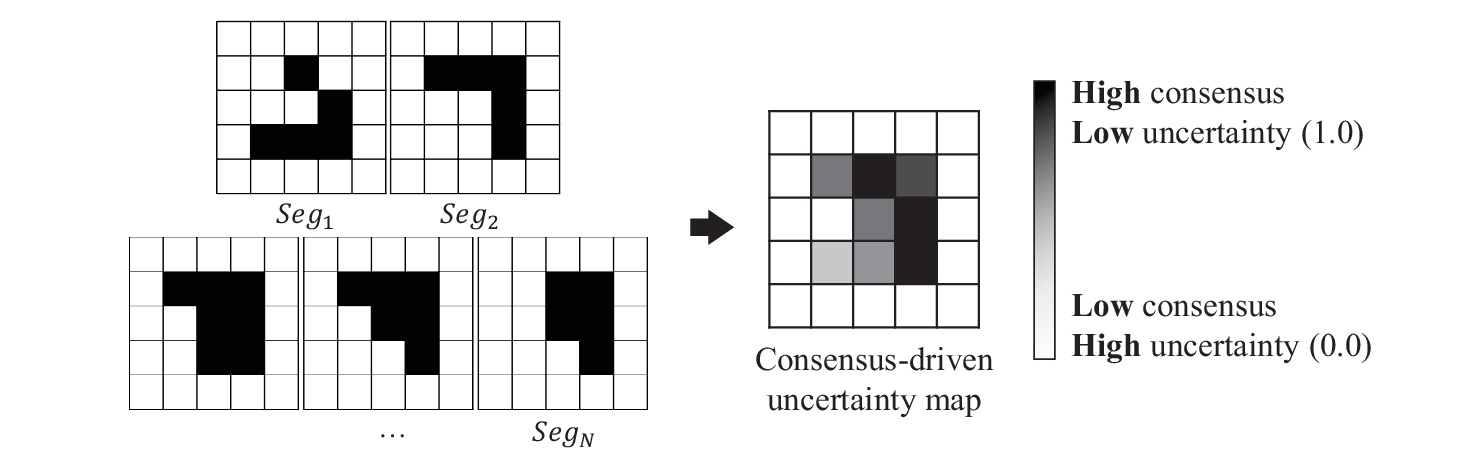}
\caption{Process of generating a consensus-driven uncertainty map} \label{fig2}
\end{figure}

To capture the level of consensus among plausible segmentations, we generate a consensus-driven uncertainty map by aggregating multiple segmentations and converting them to a probability map, as shown in Fig.~\ref{fig2}. This map provides a spatial representation of uncertainty, with higher values indicating stronger consensus level across segmentations and lower values highlighting regions of uncertainty. This uncertainty information serves as a foundation for our subsequent consensus-driven correction method.

\subsection{Consensus-driven Correction}
We are dealing with ambiguous regions that are inherently unclear in CT images, so low consensus (high uncertainty) exhibits in the consensus-driven uncertainty map. Even with low consensus, nearby pixels with similar contextual features are identified as potential pixels for correction. We leverage the consensus-driven uncertainty map, which provides valuable insight into potentially correct segmentation areas.

Consensus-driven correction minimizes the energy function in Eq. 1, which combines the unary potential (Eq. 2) and pairwise potential (Eq. 3). The unary potential is derived from the consensus level from the map, providing a pixelwise measure of segmentation confidence. The pairwise potential incorporates context information from the map, including position proximity, consensus level similarity, and gradient direction similarity. Position proximity helps maintain consistent segmentation by considering close regions, which is important for preserving anatomical continuity in thin structures. Consensus level similarity ensures that similar uncertain regions are corrected consistently. Gradient direction similarity helps maintain structural directional patterns, which is critical for thin structure to preserve shape. For example, even when consensus is low in a specific region, our method may correct a pixel classification based on its proximity to a neighboring pixel that is segmented as bone and has similar consensus level and gradient direction. The energy function is formulated as follows:

\begin{equation}
    E(x)=\sum_i{\psi_i(x_i) + \sum_{ij}{\psi_{ij}(x_i, x_j)}}
\end{equation}
\begin{equation}
    \psi_i(x_i)=-log P(x_i)
\end{equation}
\begin{align}
    \psi_{ij}(x_i, x_j) = \mu(x_i,x_j) &[w_1 \exp(-\frac{|p_i-p_j|^2}{2\theta^2_\alpha}-\frac{|c_i-c_j|^2}{2\theta^2_\beta}-\frac{|d_i-d_j|^2}{2\theta^2_\gamma}) \nonumber \\
    &+ w_2 \exp(-\frac{|p_i-p_j|^2}{2\theta^2_\delta})]
\end{align}

\noindent 
where $x_i,x_j$ represent the class of pixel $i$,$j$. $P(x_i)$ in the unary potential denotes the probability of pixel $i$ belongs to a specific class, as derived from the consensus-driven uncertainty map. The pairwise potential contains three Gaussian kernels accounting for positional proximity ($p$), consensus level similarity ($c$), and gradient direction similarity ($d$). The class compatibility function $\mu(x_i, x_j)$ is defined based on the Potts model: $\mu=1$ if $x_i\neq x_j$, $\mu=0$ otherwise. This penalizes assigning different classes to nearby, similar pixels. The parameters $\theta_\alpha, \theta_\beta, \theta_\gamma$, and $\theta_\delta$ are scaling factors for the Gaussian kernels, controlling the influence range of each characteristics. Based on experimental results and considering the image size, intensity range, and gradient direction range, we set $\theta_\alpha=80,\theta_\beta=60, \theta_\gamma=2, \theta_\delta=3, w_1=15, w_2=1$.

\section{Experiment}
\subsection{Dataset and Preprocessing}
This study was approved by the Institutional Review Board of Severance Hospital, Yonsei University College of Medicine, Seoul, Republic of Korea (IRB No. 4-2016-0603). The dataset comprises facial CT images from 71 patients, divided into a training set of 57 cases and a test set of 14 cases. All images have a matrix size of 512×512 pixels, with in-plane resolution ranging from 0.4 to 0.619mm and a slice thickness of 1mm. Three annotators—a neurosurgeon with 15 years of experience and two senior medical students—manually annotated each image according to the same annotation protocols.

We performed preprocessing to ensure consistency across different CT scanners and acquisition protocols. This included intensity normalization using a window width of 600 HU and window level of 100 HU, followed by conversion from 12-bit to 8-bit representation (0-255 intensity range). Additionally, all images were resampled to a uniform pixel spacing of 0.4 x 0.4 mm$^2$, corresponding to the highest resolution present in the dataset.

\subsection{Experimental Setup and Implementation Details}
\subsubsection{Comparison Methods.} We compare our method with CNN-based segmentation, including U-Net~\cite{ronneberger2015unet} and MSDA-Net~\cite{an2023orbital}, and diffusion-based segmentation, including MedSegDiff-v2 (Gaussian noise)~\cite{wu2024medsegdiffv2} and BerDiff (Bernoulli noise)~\cite{chen2023berdiff}.
\subsubsection{Evaluation Metrics.} Three metrics are used for performance evaluation, including Dice Similarity Coefficient (DSC), recall, and precision. As reference standard, we use masks generated by the STAPLE algorithm~\cite{warfield2004staple} from three annotations, which is widely used approach to combine multiple expert annotations to generate a reference standard. To focus on thin bone regions, we manually define two evaluation ROIs for orbital medial wall and orbital floor.
\subsubsection{Implementation Details.} The diffusion model is trained with a batch size of 2, using the AdamW optimizer with a learning rate of 5e-5 and a linear noise schedule over 1000 timesteps. The diffusion model is trained for 100,000 iterations on a NVIDIA GeForce RTX 3090 GPU using Python 3.7 and PyTorch 1.11. During training, one of multiple annotations was randomly selected for each iteration, allowing the model to learn various annotation patterns. We use DDIM sampling strategy for faster generation while maintaining quality, and run the model 200 times to generate multiple segmentations.

\subsection{Results}
\begin{figure}[t]
\includegraphics[width=\textwidth]{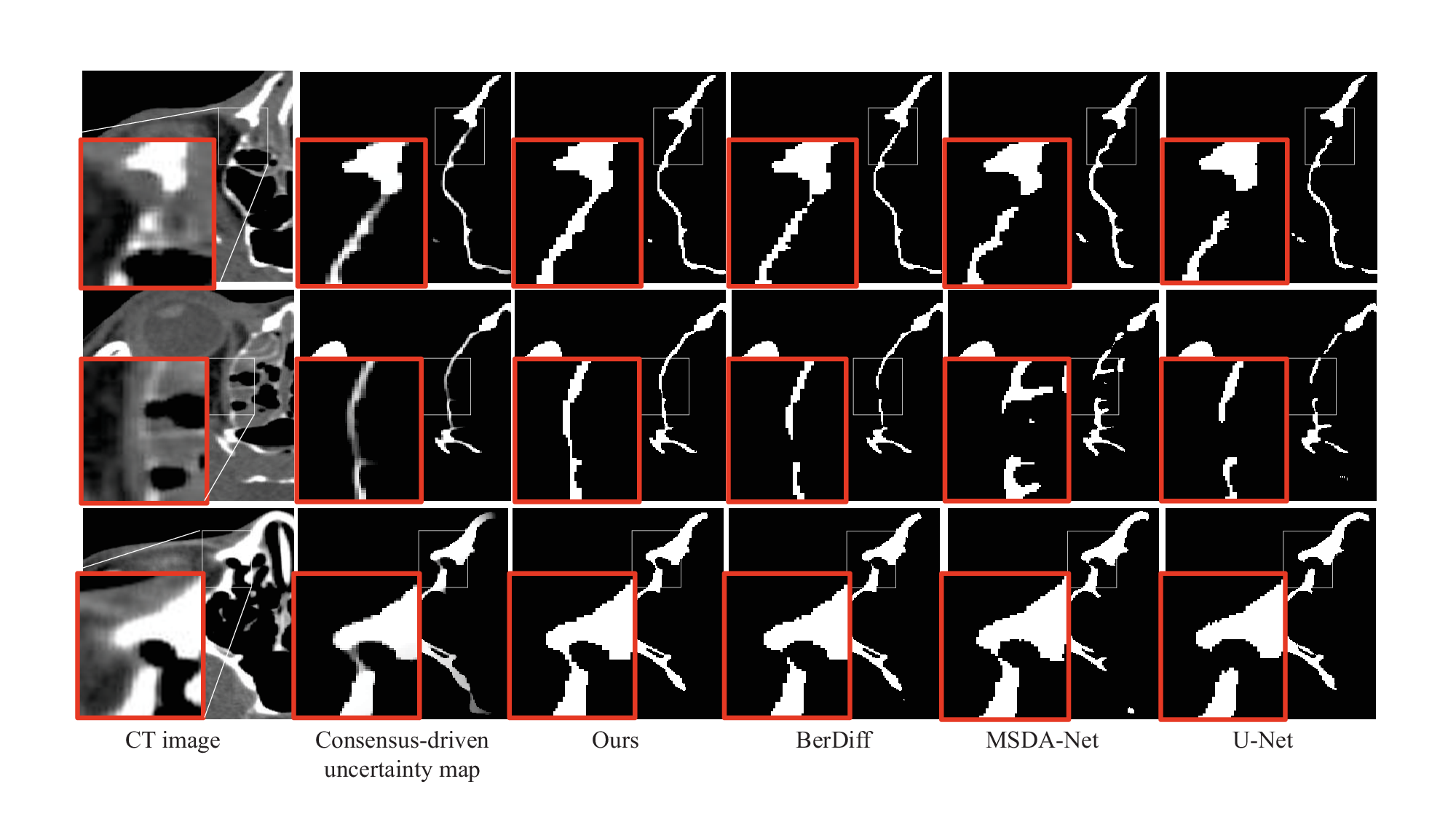}
\caption{Representative orbital bone segmentation results. Consensus-driven uncertainty map is generated by accumulating multiple segmentations using a conditional Bernoulli diffusion model (BerDiff). ‘BerDiff’ in 4th column is the mean of multiple segmentations. The red box is zoomed in to emphasize the thin region within the white box.} \label{fig3}
\end{figure}

We present the quantitative and qualitative results in Table~\ref{tab1} and Fig.~\ref{fig3}, respectively. As can be seen in Table~\ref{tab1}, our method outperforms all CNN-based methods in all metrics, including DSC, recall, and precision. In particular, the recall of the orbital medial wall showed statistically significant improvement with $p<0.001$, which effectively addressed the under-segmentation issue in thin bone regions.

\begin{table}[t]
\caption{Segmentation performance of our proposed method and comparison methods. The best results are highlighted in bold. ‘*’: p<0.001 compared to our method based on a t-test.}\label{tab1}
\begin{tabular*}{\textwidth}{@{\extracolsep{\fill}}cc|ccc|ccc}
\hline
\multicolumn{2}{c|}{\multirow{2}{*}{Methods}}                                                                             & \multicolumn{3}{c|}{Orbital medial wall}          & \multicolumn{3}{c}{Orbital floor}                \\
\multicolumn{2}{c|}{}                                                                                                     & DSC            & Recall         & Precision       & DSC            & Recall         & Precision      \\ \hline
\multicolumn{1}{c|}{\multirow{2}{*}{\begin{tabular}[c]{@{}c@{}}CNN\\ -based\end{tabular}}}       & U-Net~\cite{ronneberger2015unet}         & 82.28*         & 79.70*         & 85.92           & 89.86          & 90.24*         & 89.85          \\
\multicolumn{1}{c|}{}                                                                            & MSDA-Net~\cite{an2023orbital}      & 83.39*         & 84.18*         & 83.26*          & 89.99          & 92.72          & 87.79*         \\ \hline
\multicolumn{1}{c|}{\multirow{3}{*}{\begin{tabular}[c]{@{}c@{}}Diffusion\\ -based\end{tabular}}} & MedSegDiff-v2~\cite{wu2024medsegdiffv2} & 84.07*         & 81.58*         & 87.26*          & 90.11*         & 88.88*         & 92.02*         \\
\multicolumn{1}{c|}{}                                                                            & BerDiff~\cite{chen2023berdiff}       & 86.14          & 84.60*         & \textbf{88.16*} & 91.53          & 92.15          & \textbf{91.37} \\
\multicolumn{1}{c|}{}                                                                            & Ours                   & \textbf{86.31} & \textbf{87.83} & 85.38           & \textbf{91.36} & \textbf{93.24} & 90.08          \\ \hline
\end{tabular*}
\end{table}

Fig.~\ref{fig3} demonstrates our correction capabilities in the challenging regions. The red boxes highlight ambiguous regions, which show high uncertainty in the consensus-driven uncertainty map. CNN-based methods show disconnected segmentation in those regions. BerDiff shows improved results compared to CNN-based methods, but still fails, leaving disconnected regions. However, our proposed method successfully corrects thin structures in these challenging areas by leveraging the clear vertical directionality of the orbital medial wall through a consensus-driven correction approach.

\section{Conclusion}
In this paper, we proposed a framework combining conditional Bernoulli diffusion model-based segmentation with consensus-driven correction. This approach aims to improve the segmentation performance of thin structures in orbital bones, with a particular focus on the orbital medial wall and orbital floor, which possess ambiguous boundaries. In our experiments, we addressed under-segmentation issues in ambiguous regions by leveraging the ability of our proposed method to learn from various annotation patterns. This approach reduces dependency on subjective annotation by any single annotator and enables us to use the consensus among multiple segmentations from the diffusion model as valuable information. By considering directionality in our consensus-driven correction method, we achieved improvements of up to 4.03\% in DSC and 8.13\% in recall for orbital medial wall compared to CNN-based methods. Our novel framework learns data distribution from multiple annotations in ambiguous regions and ensures consistent segmentation results while reducing manual correction process, making it applicable to surgical planning and customized implant creation.

\begin{credits}
\subsubsection{\ackname} This research was partially supported by a grant of the Korea Health Technology R\&D Project through the Korea Health Industry Development Institute (KHIDI), funded by the Ministry of Health \& Welfare, Republic of Korea (grant number: HI22C1496) and the National Research Foundation of Korea (NRF) grants funded by the Korea government (MSIT) (RS-2024-00340610 and RS-2023-00207947).
\subsubsection{\discintname}
The authors have no competing interests to declare that are relevant to the content of this article.
\end{credits}

%
%
%
\bibliographystyle{splncs04}
\bibliography{Paper-5352}
\end{document}